\newcommand{\nat}[2]{Nat #1, #2}
\newcommand{\apj}[2]{ApJ #1, #2}
\newcommand{\apjs}[2]{ApJS #1, #2}

\newcommand{\aj}[2]{AJ #1, #2}

\newcommand{\aeta}[2]{A\&A #1, #2}

\newcommand{\aetas}[2]{A\&AS #1, #2}

\newcommand{\mn}[2]{MNRAS #1, #2}

\newcommand{\nh}{N$_{\rm H}$}

\newcommand{\kms}{km s$^{-1}$}

\newcommand{\Tbb}{$\rm T_{\rm bb}$\,}

\newcommand{\degree}{\degr}

\newcommand{\Lx}{$\rm L_{\rm X}$}

\def\refb{\par\noindent\hangindent 15pt}
\documentstyle[bo99,epsfig]{article}

\title{Isolated neutron stars discovered by ROSAT}

\author{C.~Motch $^1$}                                                       
\affil{1) CNRS, Observatoire de Strasbourg, 11 rue de l'Universit\'e, 
67000 Strasbourg, France}                                                
\begin{document}

\maketitle

\begin{abstract}

ROSAT has discovered a new group of isolated neutron stars characterized by soft
black-body like spectra (kT $\sim$ 50-120\,eV), apparent absence of radio emission and
no association with supernovae remnants. So far only six such sources are known. A
small fraction of these stars exhibit X-ray pulsations with relatively long periods of
the order of 10 sec. Two very different mechanisms may be envisaged to explain their
properties. The neutron stars may be old and re-heated by accretion from the ISM in
which case their population properties could provide information on past stellar
formation and secular magnetic field decay. Alternatively, this group may at least
partly  be made of relatively young cooling neutron stars possibly descendant from
magnetars. We review the last observational results and show how they can shed light on
the evolutionary path of these new objects within the whole class of isolated neutron
stars.

\keywords{neutron stars; pulsars; magnetars}               
\end{abstract}

\section{Introduction}

The vast majority of the 10$^{8}$-10$^{9}$ isolated neutron stars (INS) present in the
Galaxy should be virtually undetectable using current observational means. Young cooling
neutron stars might emit thermal X-rays during the first $\sim$ 10$^{6}$\,yr and their
pulsed radio emission will reveal them up to ages of $\sim$ 10$^{8}$\,yr. Recycled
millisecond pulsars are old objects but their previous accreting binary phase may have
altered their physical properties in particular the magnetic field strength. In this
context, the possibility that a sizeable fraction of the entire 'fossil' population is
re-heated by accretion from interstellar medium and becomes detectable in the EUV /
X-ray domain is exciting. This could allow an observational study of old neutron stars
and give access to information on past stellar formation, heating and cooling mechanisms
and magnetic field decay. This idea was first proposed by Ostriker, Rees \& Silk (1970).
Population synthesis models were later computed by Treves \& Colpi (1991), Blaes \&
Madau (1993) and Madau \& Blaes (1994). A recent and extensive review on current issues
can be found in Treves et al. (1999). Early models predicted that a rather large number
of ROSAT XRT all-sky survey (RASS) sources could be accreting INS, initiating several
optical identification campaigns. In this paper we shall describe the general
observational properties of the INS discovered by ROSAT, discuss the possible X-ray
powering mechanisms, consider what identification campaigns can tell us so far on the
total population and discuss in more details the pulsating sources.

%--------------------------------  Figure 1
\begin{figure}
\centerline{\psfig{file=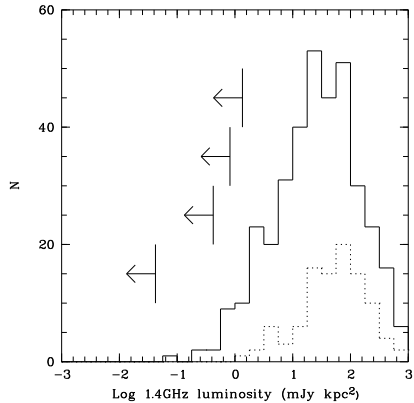, angle=-90, width=8cm}}
\label{radioul}
\caption[]{Upper limits on 1.4GHz radio luminosity of ROSAT discovered INS. Vertical
displacement is there only for readability.  Histograms represent the distribution of
observed luminosities of radio pulsars as in Taylor et al. (1993) (Total population is
the continuous line, pulsars with ages below 10$^{6}$\,yr is the dashed line)}
\end{figure}
%---------------------------------

\section{General properties}

The group of INS discovered by ROSAT shares a rather well defined set of properties.
Although each of these features may be present individually in sub groups of INS (e.g.
some radio pulsars have very soft X-ray spectra and some X-ray emitting radio quiet INS
are found in SNR environments), the properties characterizing this new group are never
encountered together in previously known classes of INS.

\underline{Soft X-ray spectra.} All ROSAT discovered INS exhibit soft X-ray spectra
which at the energy resolution of the PSPC can be very well fitted by blackbody models
with temperatures in the range of $\sim$ 50 to 120 eV. No bright X-ray hard component
seems to be present, at least at the level encountered in most X-ray detected radio pulsars.
Another common spectral feature is the low \nh \ towards these sources ($\sim$
10$^{20}$ cm$^{-2}$) which implies distances of the order of 100 to 1000\,pc at most. 

% ------------------------------- Table 1
\begin{table*}
\caption{Upper limits on radio emission from ROSAT discovered INS}
\vskip 0.5cm
\label{radiointensity}
\begin{tabular}{ccccc}
\hline
ROSAT      &Frequency   & Assumed       & Max L  & Survey \\
source     &  GHz       & distance (kpc) & mJy kpc$^{2}$  &        \\\hline
RX J1856$-$3754 & 0.43  &  0.13          & 0.058          & Parkes survey \\
               & 1.4   &                & 0.042          & NVSS    \\
RX J0720$-$3125& 1.4   &  0.41          & 0.420           & NVSS    \\
RX J1605+3249   & 1.4   &  0.70          & 0.400           & FIRST   \\ 
RX J1308+2127   & 1.4   &  1.20          & 1.350           & FIRST   \\
\hline
\end{tabular}
\end{table*}

\underline{No strong radio emission.} The absence of detected bright radio emission
seems to be also a common feature although not all members have been surveyed yet. At
present, constraints on radio emission are mostly based  on the FIRST and NVSS surveys.
We list in Table 1 various upper limits on luminosity gathered for a subset of INS
candidates. In Fig. 1 we show the position of these upper limits with respect to
observed 1.4 GHz luminosity distribution of radio pulsars.

ROSAT discovered INS are undoubtedly less radio luminous than most known
radio pulsars. The difference is stricking if one assumes that these are young
cooling objects with ages of less than about 10$^{6}$\,yrs. However, targeted
radio observations should easily improve on current upper limits. 

\underline{No association with SNR.} ROSAT and Einstein observatories have discovered a
number of X-ray bright but apparently radio quiet neutron stars associated with SNRs
(Gotthelf et al. 1997, Brazier \& Johnston 1999). Their X-ray spectra, although often
thermal like are in general significantly hotter than those of the group of INS
discussed here. In contrast, the fact that none of the INS considered here lies close
to an SNR suggests ages older than $\sim$ 10$^{5}$\,yr. 

\underline{No long term variability.} ROSAT was able to monitor the mean X-ray
luminosity of some candidates over several years, basically from the all-sky survey time
till the last operational phases. EXOSAT and Einstein have also serendipitously observed
some of these targets further extending the time base. In the cases investigated in
details no evidence for X-ray variability over months or years time scale exist with
very stringent upper limits of the order of few percents (Walter et al. 1996, Haberl et
al. 1997, Motch et al. 1999).

\underline{Long rotation periods.} At least one of the INS candidates, RX\,J0720-3125
exhibits X-ray pulsations with a period of 8.39\,s (Haberl et al. 1997). Another case
is RX J0420$-$5022 which is possibly pulsating at 22.7\,s (Haberl et al. 1999). The
spin period of other candidates is not known but this issue may be settled by
forthcoming XMM observations. These long spin periods are most unusual for radio
pulsars. We shall see below that the long periods have profound impact on the
possible evolutionary status of these objects.

\subsection{Optical identifications}

Only the two X-ray brightest candidates have a secure optical identification. The ROSAT
source RX\,J1856.5$-$3758 was identified by Walter et al. (1996) with a V = 25.6, U-V =
$-$1.2 object. A faint B = 26.1-26.5 blue star is also the likely counterpart of
RX\,J0720-3125 (Motch \& Haberl 1998, Kulkarni \& van Kerkwijk 1998). In these two
cases, the optical continuum lies only 1.4 and 1.7 magnitude respectively above the
Rayleigh-Jeans tail of the blackbody seen in soft X-rays. For the remaining candidates,
the identification with an INS eventually rests on the absence of optical signatures of
other possible soft X-ray emitters such as magnetic CVs or soft AGNs. 

\subsection{The catalogue}

We list in Table 2 the main properties of the INS discovered so far on the basis of
their X-ray emission. In almost all cases, their nature was established by ROSAT
observations although some of them already had Einstein or EXOSAT detections. MS
0317.7-6647 (Stocke et al. 1995) is a particular case as its X-ray energy distribution
undergoes very high absorption. Also, its INS nature is not entirely clear since an
identification with a very massive accreting black hole in the field spiral galaxy NGC
1313 is not excluded. 

\begin{table*}
\caption{The catalogue of isolated neutron stars discovered by their X-ray emission}
\vskip 0.5cm
\label{xintensity}
\begin{tabular}{ccccccc}
\hline
ROSAT      &PSPC  &  kT & \nh                  & P &B mag  &References \\
source     &cts/s & eV  & 10$^{20}$\,cm$^{-1}$ & s &       & \\
\hline
RX J1856$-$3754&  3.64  &  57 $\pm$ 1 & 1.4 $\pm$ 0.1 &  -  &25.8     & 1 \\
RX J0720$-$3125&  1.64  &  79 $\pm$ 4 & 1.3 $\pm$ 0.3 &8.39 &26.5     & 2 \\
RX J1605+3249&  0.90  &  92 $\pm$ 6 & 1.1 $\pm$ 0.4 &  -  & $>$25   & 3   \\
RX J0806$-$4123&  0.38  &  78 $\pm$ 7 & 2.5 $\pm$ 0.9 &  -  & $>$24   & 4 \\
RX J1308+2127&  0.29  &  113 $\pm$ 14 & 2.0 $\pm$ 0.4 &  -  & $>$26   & 5 \\
RX J0420$-$5022&  0.14  &     $<$85   &    ~ 2.0    &22.7: & $>$25.2 & 6 \\
MS 0317.7$-$6647 &$\sim$ 0.03 & 180 $\pm$ 30 & 40 $\pm$ 20 &  -  & $>$ 21.4 & 7 \\
\hline
\end{tabular}
%\end{table*}
References: 1) Walter et al. (1996), 2) Haberl et al. (1997), 3) Motch et al.
(1999), 4) Haberl et al. (1998), 5) Schwope et al. (1999), 6) Haberl et al.
(1999), 7) Stocke et al. (1995).
\end{table*}

\section{X-ray powering mechanisms}

A number of possible mechanisms leading to production of X-rays from INS may be
envisaged. Basically, the X-ray luminosity may be extracted from spin down as 'normal'
radio pulsars do, from cooling, or the neutron star may be re-heated by accretion of
interstellar material. We consider below in some details each of these mechanisms.

\subsection{Rotation} Becker \& Tr\"umper (1997) showed that for 'normal' radio
pulsars, the X-ray luminosity is tightly linked to rotational energy loss with \Lx $\sim$
10$^{-3}$ $\dot{\rm E}$. Wang et al. (1999) pointed out that at least in the case of RX
J0720$-$3125 this mechanism could not work since the observed X-ray luminosity was more
than twice the upper limit on spin down power ($\dot{\rm P}$ $\leq$ 0.8 10$^{-12}$,
Haberl et al. 1997). However in the absence of spin period measurements and derivatives
this mechanism cannot be ruled out for other INS candidates. All radio pulsars bright
enough to have a constraining ROSAT PSPC spectrum exhibit a rather hard power law
component which is thought to be the signature of intense magnetospheric activity
(Becker \& Tr\"umper 1997). The absence of a luminous hard X-ray tail above the thermal
component in ROSAT discovered INS is a general argument against a rotationally driven
origin of X-rays.

\subsection{Cooling} Blackbody temperatures in excess of 50\,eV imply ages younger than
$\sim$ 10$^{6}$\,yr for standard cooling curves while the absence of nearby SNR sets a
lower limit of $\sim$ 10$^{5}$\,yr. Among the X-ray emitting radio pulsars listed by
Becker \& Tr\"umper (1997) two middle aged (10$^{5}$-10$^{6}$\,yr) pulsars (PSR
0656+14 and PSR 1055-52) exhibit in addition to the power law a clear thermal
component believed to be due to cooling from the neutron star surface. A luminous
thermal component is probably also present in the younger (10$^{4}$-10$^{5}$\,yr) Vela
type pulsars. In PSR 0656+14 the soft blackbody dominates the 0.1-2.4 keV energy
distribution and has characteristics (\Tbb = 87\,eV, \nh \ = 1.9 $\pm$ 0.4 10$^{20}$
cm$^{-2}$) quite similar to those of the INS discussed here. In the case of RX
J1605+3249 for instance, Motch et al. (1999) show that owing to the lower statistics, a
hard component of similar intensity as in PSR 0656+14 would not have been detected nor
would the 0.384\,s period with a 9\% pulsed fraction have been seen. In this context, a
natural and simple explanation for the absence of radio emission from an otherwise
X-ray bright and nearby source is that the radio beam does not sweep the earth. The
proportion of the sky swept by the radio beam decreases with increasing periods and is
of the order of 0.3 for the period range of cooling pulsars (Biggs 1990). In the range
of PSPC count rates covered so far there are two cooling radio pulsars (see Table 3).
Although this is small number statistics, the picture is in fact consistent with the
entire ROSAT discovered population being cooling radio pulsars whose beam remains
undetected because it does not cross the earth.  

\begin{table}
\caption{Brightest X-ray pulsars in the ROSAT PSPC instrumental system after
Becker \& Tr\"umper (1997)}
\vskip 0.5cm
\begin{tabular}{ccccccc}
\hline
Pulsar      &PSPC  & P    & Log(P/2$\dot{\rm P}$)  &Note\\
name        &cts/s & ms   & years         & \\
\hline
Crab        & 17.8 & 33.4 & 3.10     &  \\
Vela        & 3.40 & 89.29& 4.05     &  \\
B0656+14    & 1.92 & 384.87 & 5.05   & cooling pulsar \\
B1055-52    & 0.35 & 197.10 & 5.73   & cooling pulsar \\
J0437-47    & 0.20 & 5.75   & 9.50   & \\ 
\hline
\end{tabular}
\end{table}

\subsection{Accretion from interstellar medium} Accretion of interstellar matter onto
old neutron stars can produce X-ray luminosities large enough to detect a sizeable
fraction of the total galactic population. Assuming Bondi-Hoyle accretion, the mass
accretion rate is $\dot{\rm M}$ $\sim$ 10$^{10}$ n $({\rm V}/40{\rm\, kms^{-1}})^{-3}$
g\,s$^{-1}$ with V the neutron star velocity  with respect to interstellar medium and n
the mean density. The temperature of the polar cap heated by accretion is \Tbb \ $\sim$
20 ($\dot{\rm M}$/10$^{10}$g\,s$^{-1}$)$^{1/4}$ f$^{-1/4}$ eV with f the relative
surface area of the polar cap. However, before matter can reach the neutron star surface
a number of conditions must be fulfilled. The first one is that ram pressure at
accretion radius exceeds pulsar momentum flux otherwise the star is in the ejector
state. This requires the spin period to be less than $\sim$ 9 $(\rm B/10^{12}\rm
G)^{1/2}$ n$^{-1/4}$ $({\rm V}/40{\rm\, kms^{-1}})^{1/2}$\,s. Magnetic dipole braking
allows to reach this period in $\sim$ 5 $(\rm B/10^{12} \rm G)^{-1}$ n$^{-1/2}$ $({\rm
V}/40{\rm\, kms^{-1}})$ Gyr (Blaes \& Madau 1993). The most constraining condition is
that the corotation radius must be larger than the Alfven radius. For the low accretion
rates prevailing here, this condition requires very slow rotation P$_{\rm rot}$ $\geq$
3000 $({\rm B}/10^{12})^{6/7}$ n$^{-3/7}$ $({\rm V}/40{\rm\, kms^{-1}})^{9/7}$\,s.
Dipole magnetic radiation is not efficient enough to slow down the neutron star on a
reasonable time scale. Braking is probably achieved by the propeller mechanism which
acts on a short time scale of $\sim$ 0.8 $({\rm B}/10^{12}\rm G)^{-11/14}$ n$^{-17/28}$
$({\rm V}/40{\rm \, kms^{-1}})^{29/14}$ Gyr (Blaes \& Madau 1993). 

\section{Constraints on populations of X-ray emitting isolated neutron stars} The first
constraints came from the general identification campaigns which aimed at a complete
census of the RASS source population at different flux levels and galactic latitudes.
In addition a number of projects specifically searching for INS were initiated.
Zickgraf et al. (1997), Motch et al. (1997), Belloni et al. (1997) and Danner et al.
(1998) reported on deep surveys in small areas including molecular clouds. Shallow
surveys of large areas were performed by Manning et al. (1996), Bade et al. (1998) and
Thomas et al. (1998), among others. These campaigns led to the discovery of the handful
of candidates discussed here. They also yielded the unexpected result that the number
of INS present in the RASS was far below that predicted by early population synthesis
models which assumed that a few thousands neutron stars should be detected in the RASS.
The results of these observational constraints have been summarized by Neuh\"auser \&
Tr\"umper (1998). We show an updated version of their figure 1 in Fig. 2. At the faint
flux end, upper limits on ROSAT INS densities are incompatible with model predictions
from Treves \& Colpi (1991) and Blaes \& Madau (1993) (not shown) but still consistent
with the revised models of Madau \& Blaes (1994). The most recent large sky area
searches (labeled A,B and C in Fig. 2) yield even stronger constraints. The results
(A) of Danner et al. (1998) is restricted to dark clouds area and is therefore not
directly comparable with model predictions. Constraints labeled B and C cover large
unbiased sky areas. (C) results from the identification of very bright soft ROSAT
sources with $|b|$ $\geq$ 20\degree \ by Thomas et al. (1998). The upper limit shown in
(B) is derived from an identification campaign of soft sources over $\sim$ 1/4 of the
whole sky conducted at ESO (Motch et al. 2000). Optical identification of a total of
116 sources down to a PSPC count rate of 0.17 cts/s yielded only 3 possible cases among
which 2 are already known INS. Spectral selection is discussed in Motch et al. (1999).

Contrary to other studies, surveys (B) and (C) assume that accreting INS are
intrinsically soft sources and preselect candidates on the basis of ROSAT PSPC hardness
ratios. The brightest INS found in the RASS without spectral selection are indeed soft
and a soft spectrum is also theoretically expected (e.g. Zampieri et al. 1995).
However, the possibility remains that because of the spectral preselection a fraction
of the population, for instance INS accreting in very dense medium, escapes the search.

The painstaking optical identification campaigns have yielded the very significant and
unexpected result that the number of accreting old INS shining in X-rays is a factor 10
to 1000 below that predicted by population synthesis models. This discrepancy points to
a fundamental error in one or more of the model assumptions used. Constraints on an
accreting population may be even stronger since; i) the observed LogN-LogS is almost
consistent with a young cooling neutron star population only and ii) for none of the
INS candidates discovered so far can the accreting model be unambiguously established
by the discovery of X-ray variability or the measurement of a negative $\dot{\rm P}$
for instance.  In two cases INS are found in regions of particularly low mean densities
contrary to naive expectations (n $\leq$ 0.4 and 0.3 cm$^{-3}$ for RX J0720$-$3125 and
RX J1605+3249 respectively). Very low relative velocities of less than 10\,\kms \ must
be assumed to account for the observed luminosities. However, selection effects and ISM
patchiness on very small scales may solve this paradox. Finally, accretion should
enrich neutron star atmosphere in H and He and could cause a brighter optical continuum
than observed (Pavlov et al. 1996).

%--------------------------------  Figure 2
\begin{figure}
\centerline{\psfig{file=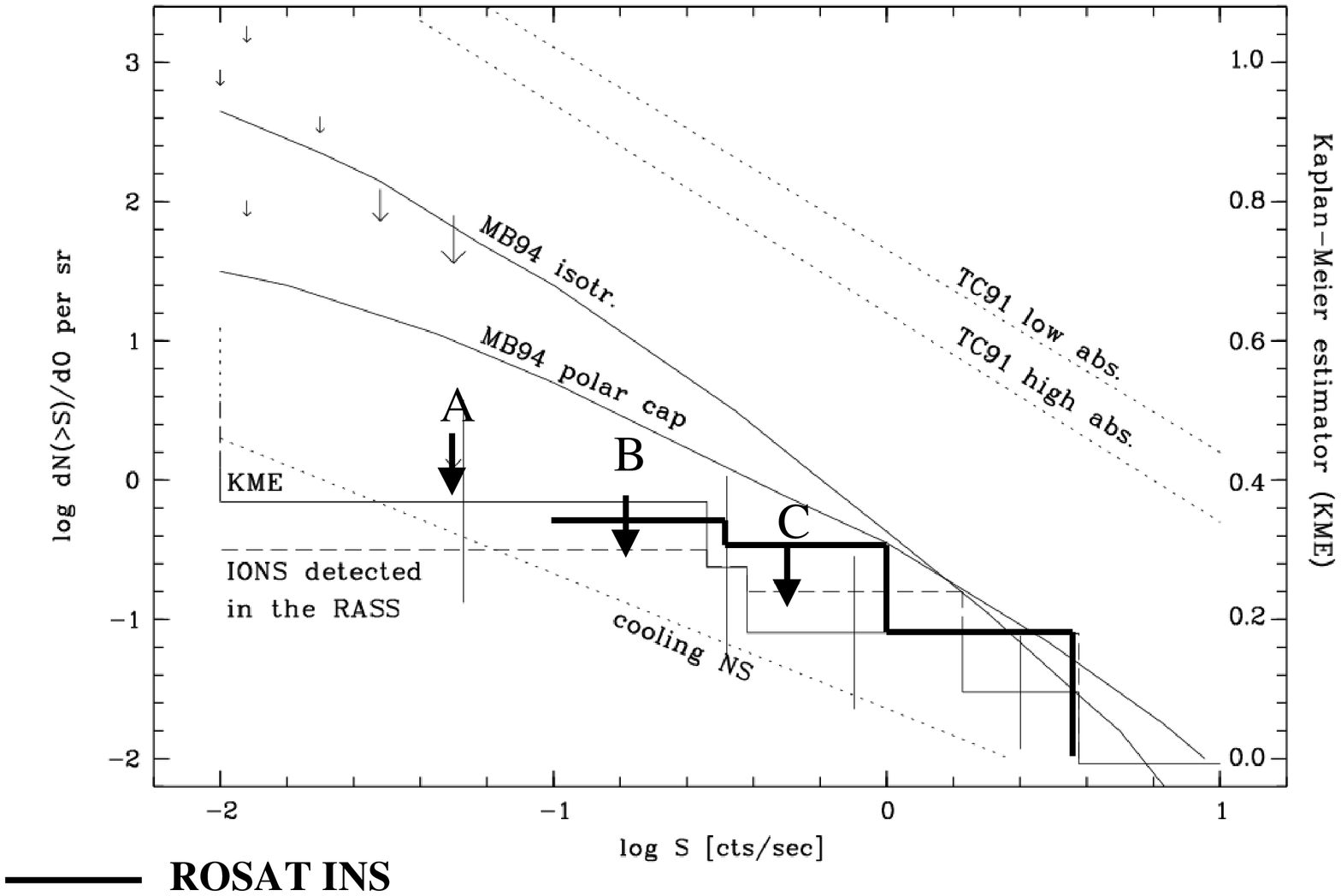,bbllx=3.0cm,bburx=28cm,bblly=2.0cm,bbury=18cm,width=9cm, clip=true}}
\caption[]{LogN-LogS curves for X-ray detected isolated neutron stars after 
Neuh\"auser \& Tr\"umper (1998). Lines labelled TC91 amd MB94 represent model
predictions from Treves \& Colpi (1991) and Madau \& Blaes (1994). Thin arrows are
upper limits derived from various identification campaigns. The thick curve is the
revised INS LogN-LogS curve. Thick arrows marked A,B and C represent upper limits
resulting from A) dark clouds (Danner et al. 1998), B) Soft sources on 1/4 of the
sky (see text), C) RASS soft sources (Thomas et al. 1998).}
\end{figure}
%---------------------------------

\section{Are old neutron stars accreting ?} Newly born neutron stars must be spun down
efficiently before they can start accreting from the interstellar medium. The evolution
of magnetic field with time seems a key parameter as it both determines braking strength
during ejector and propeller phases and the minimum period allowing accretion. Treves et
al. (1999) argue that for certain values of initial magnetic field and decay time
scales, the life time of the ejector phase may be longer than the age of the Galaxy.
Colpi et al. (1998) and Livio et al. (1998) also showed that field decay significantly
increases the duration of the propeller phase with the consequence that most old neutron
stars could still not be able to accrete. Because of the steep dependence of Bondi-Hoyle
accretion rate with relative velocity, the velocity distribution of old neutron stars is
another key parameter. Early models were based on the work of Narayan \& Ostriker
(1990). Since then new observations by Lyne \& Lorimer (1994) suggest a typical mean
velocity at birth of the order of 450\, \kms \ about twice that of Narayan \& Ostriker
(1990). These two effects may cooperate to explain the small number of X-ray detections.

\section{Pulsating sources} Two INS, RX J0720$-$3125 and possibly RX J0420$-$5022 have
long rotation periods which modulate their X-ray light curve by 20 to 80 \%. These spin
period are significantly longer than those generally measured in radio pulsars. In the
accretion scenario, the spin period yields an estimate of the surface magnetic field
which has to be weaker than $\sim$ 10$^{10}$\,G to allow for accretion (Haberl et al.
1997). If confirmed, this would clearly point to secular magnetic field decay.
Unfortunately, alternative explanations exist. Based on the similarity of spin periods
Haberl et al. (1997) proposed that RX J0720$-$3125 is related to class of anomalous
(braking) X-ray pulsars (AXPs). The nature of AXPs itself remains a mystery. Two models
often mentioned are i) isolated neutron stars evolving from Thorne-$\dot{\rm Z}$ytkow
objects (van Paradijs et al. 1995) and accreting from a remnant disk and ii) magnetars
(Thompson \& Duncan 1996). An interpretation in terms of a cooling neutron star faces
the difficulty to reconcile thermal and rotational ages. For RX J0720$-$3125, \Tbb = 8
10$^{5}$ K implies ages in the range of 1 to 4 10$^{5}$\,yr (Heyl \& Hernquist 1998)
whereas the time required to spin down from a short birth period to 8.39\,s with dipole
radiation is $\sim$ 1.2 10$^{9}$ $(\rm B/10^{12}\rm G)^{-2}$\,yr. This discrepancy may
be solved assuming the star was born with a long period. Alternatively, Heyl \&
Hernquist (1998) proposed the exciting possibility that RX J0720$-$3125 is a magnetar
with B$\sim$ 10$^{14}$ G, similar to those proposed to explain soft-$\gamma$ repeaters
and AXPs. Because radio emission is quenched by the strong magnetic field magnetars
remain undetected by classical radio means. Their birth rate may be $\sim$ 10\% of that
of ordinary pulsars (Kouveliotou et al. 1994). Gotthelf \& Vasisht (1999) also argue
that a large fraction of SNR contains previously unrecognized slowly rotating,
radio-quiet and X-ray bright pulsars which could be the natural progenitors of the long
period INS discovered by ROSAT. The high magnetic field allows efficient magnetic dipole
spin down and its decay provides an additional source of heat which allows magnetars to
remain detectable in X-rays over longer times than ordinary pulsars. Consequently,
although less numerous than normal pulsars, cooling magnetars may show up in comparable
numbers in X-ray surveys.

\section{Conclusions} All optical identification campaigns carried out so far confirm
the scarcity of X-ray emitting isolated neutron stars in the ROSAT all sky survey.
Clearly, the large population of old galactic neutron stars does not
accrete from interstellar medium at the rate predicted by early models, or the
accretion energy is re-radiated outside the ROSAT energy band. This unexpected result
tells us something fundamental on the kinematics and/or magnetic field evolution of old
neutron stars and points to a secular magnetic field decay or a large mean velocity.
ROSAT has however discovered a handful of INS sharing as common properties a soft
blackbody like X-ray spectrum, absence of luminous radio emission and absence of long
term variability. It is not yet proven that any of  the 6-7 confirmed INS accretes from
ISM. Their small number could be compatible with cooling normal pulsars whose radio
beam does not cross the earth or with magnetars evolved from Soft-$\gamma$ repeaters,
AXPs and more generally from the slowly rotating X-ray bright and radio quiet pulsars
found in several SNRs. Future observations with VLTs and Chandra/XMM should help to solve
the puzzle of the evolutionary status of this population by measuring proper motions,
rotation periods and period derivatives which are key parameters to distinguish between
young cooling and old accreting INS. Finally these INS should allow the detailed
observation of a rather unperturbed neutron star atmosphere and the accurate
measurement of fundamental neutron star properties.

\begin{acknowledgements}
I would like to thank F. Haberl for enlightening discussions.
\end{acknowledgements}

\end{document}